%% file: paper.tex
\documentclass[acmtog]{acmart}
\usepackage[ruled,vlined]{algorithm2e}
\usepackage{hyperref}

\usepackage{bbding}
\RestyleAlgo{ruled}
\RestyleAlgo{vlined}

\acmJournal{TOG}

\usepackage{booktabs}

\input{macros}
\input{macro} % macros Beibei
\usepackage{graphicx}
\usepackage{subcaption}
\usepackage{multirow}
\usepackage{makecell, multicol, diagbox}
\newcommand{\tabincell}[2]{\begin{tabular}{@{}#1@{}}#2\end{tabular}}

\makeatletter
\newcommand\paragraphNew{\@startsection{paragraph}{4}{\parindent}%
  {-.5\baselineskip \@plus -2\p@ \@minus -.2\p@}%
  {-3.5\p@}%
  {\ACM@NRadjust{\@parfont}}}
	\makeatother
	
\AtBeginDocument{%
  \providecommand\BibTeX{{%
    \normalfont B\kern-0.5em{\scshape i\kern-0.25em b}\kern-0.8em\TeX}}}

\acmJournal{TOG}

\begin{document}


\title{TensoSDF: Roughness-aware Tensorial Representation for Robust Geometry and Material Reconstruction}

\author{Jia Li}
\affiliation{
 \institution{School of Software, Shandong University}
 \country{China}
}
\email{riga27527@gmail.com}

\author{Lu Wang$^\dagger$}
% \authornote{Corresponding author.}
\thanks{$^\dagger$Corresponding authors.}
\affiliation{
  \institution{School of Software, Shandong University}
  \country{China}
}
\email{luwang_hcivr@sdu.edu.cn}

\author{Lei Zhang}
\affiliation{
  \institution{The Hong Kong Polytechnic University}
  \country{China}
}
\email{cslzhang@comp.polyu.edu.hk}

\author{Beibei Wang$^\dagger$}
% \thanks{$^\dagger$Corresponding author.}
\affiliation{
   \institution{School of Intelligence Science and Technology, Nanjing University}
   \country{China}
}
\email{beibei.wang@nju.edu.cn}

% \renewcommand{\shortauthors}{Cui et al.}

\begin{abstract}
Reconstructing objects with realistic materials from multi-view images is problematic, since it is highly ill-posed. Although the neural reconstruction approaches have exhibited impressive reconstruction ability, they are designed for objects with specific materials (e.g., diffuse or specular materials). To this end, we propose a novel framework for robust geometry and material reconstruction, where the geometry is expressed with the implicit signed distance field (SDF) encoded by a tensorial representation, namely \emph{TensoSDF}. At the core of our method is the roughness-aware incorporation of the radiance and reflectance fields, which enables a robust reconstruction of objects with arbitrary reflective materials. Furthermore, the tensorial representation enhances geometry details in the reconstructed surface and reduces the training time. Finally, we estimate the materials using an explicit mesh for efficient intersection computation and an implicit SDF for accurate representation. Consequently, our method can achieve more robust geometry reconstruction, outperform the previous works in terms of relighting quality, and reduce 50\% training times and 70\% inference time. Codes and datasets are available at \href{https://github.com/Riga2/TensoSDF}{https://github.com/Riga2/TensoSDF}.
\end{abstract}

\setcopyright{acmlicensed}
\acmJournal{TOG}
\acmYear{2024} \acmVolume{43} \acmNumber{4} \acmArticle{150} \acmMonth{7}\acmDOI{10.1145/3658211}


% \ccsdesc[500]{Computing methodologies~Reflectance modeling}

\begin{teaserfigure}
\centering
\includegraphics[width=\textwidth]{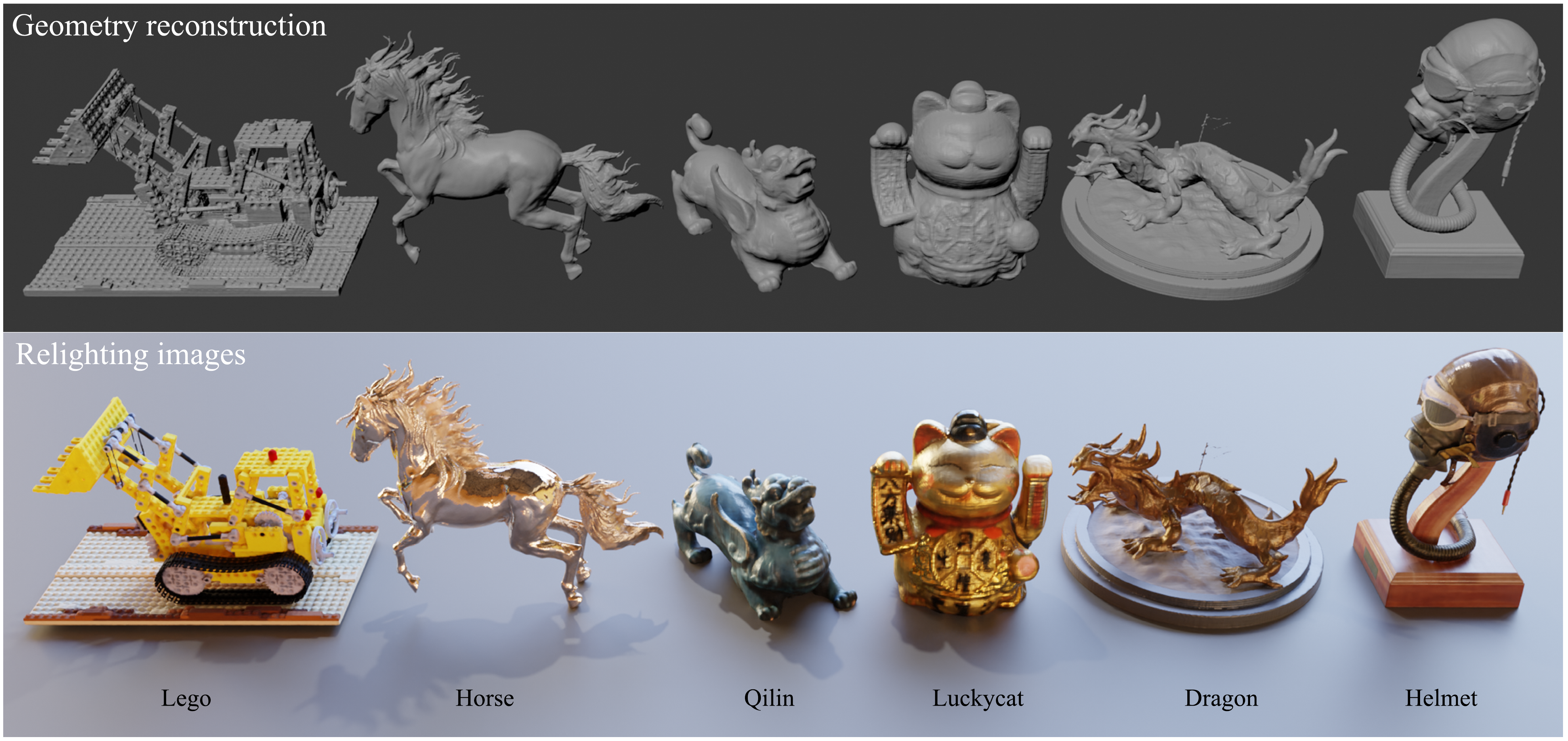}
\caption{We present a roughness-aware tensorial representation for robust geometry and material reconstruction from multi-view images. In this scene, we showcase six objects with different materials, including diffuse \textsc{Lego} from TensoIR~\cite{jin2023tensoir}, specular \textsc{Horse} from NeRO~\cite{liu2023nero}, and several glossy objects from NeILF++~\cite{zhang2023neilf++} and our datasets, where \textsc{Qilin} and \textsc{Luckycat} are real data. Our method demonstrates robust reconstruction of any reflective objects, detailed geometry results and faithful material estimation, leading to photo-realistic relighting.
}
\label{fig:teaser}
\end{teaserfigure}

\maketitle

\input{sec_1-intro}
\input{sec_2-related}

\input{sec_3-method}

\input{sec_4-implementation}

\input{sec_5-results}
\input{sec_6-conclusion}

\begin{acks}
We thank the reviewers for the valuable comments, and XiaoLong Wu for his help in our video. This work has been partially supported by the National Science and Technology Major Project under grant No. 2022ZD0116305 and National Natural Science Foundation of China under grant No. 62272275 and 62172220. 
\end{acks}

%\newpage 
\bibliographystyle{ACM-Reference-Format}
\bibliography{paper}

\end{document}


\title{Supplementary Material -- TensoSDF: Roughness-aware Tensorial Representation for Robust Geometry and Material Reconstruction}

\author{Jia Li}
\affiliation{
 \institution{School of Software, Shandong University}
 \country{China}
}
\email{riga27527@gmail.com}

\author{Lu Wang$^\dagger$}
% \authornote{Corresponding author.}
\thanks{$^\dagger$Corresponding authors.}
\affiliation{
  \institution{School of Software, Shandong University}
  \country{China}
}
\email{luwang_hcivr@sdu.edu.cn}

\author{Lei Zhang}
\affiliation{
  \institution{The Hong Kong Polytechnic University}
  \country{China}
}
\email{cslzhang@comp.polyu.edu.hk}

\author{Beibei Wang$^\dagger$}
% \thanks{$^\dagger$Corresponding author.}
\affiliation{
   \institution{School of Intelligence Science and Technology, Nanjing University}
   \country{China}
}
\email{beibei.wang@nju.edu.cn}

% \renewcommand{\shortauthors}{Cui et al.}

\maketitle

\input{supp_content}

%\begin{acks}
 %   We thank the reviewers for the valuable comments. This work has been partially supported by the National Key R\&D Program of China under grant No. 2022ZD0116305 and National Natural Science Foundation of China under grant No. 62172220.
%\end{acks}

% \begin{acks}
% \end{acks}

%\newpage 
\bibliographystyle{ACM-Reference-Format}
\bibliography{paper}

% Appendix
% \appendix
% \input{paper_appendix}

%% file: macros.tex
\SetKwComment{Comment}{// }{}

%% file: macro.tex
\def\figurePath{fig/}

\def\mycfigure#1#2#3{\begin{figure*}[htb]\centering\includegraphics*[clip, width = \linewidth]{\figurePath#2}\caption{#3}\label{fig:#1}\end{figure*}}

\def\myfigures#1#2#3#4{\begin{figure}[htb]\centering\includegraphics*[width = #4]{\figurePath#2}\caption{#3}\label{fig:#1}\end{figure}}

%% file: sec_1-intro.tex
\section{Introduction}

Reconstructing objects with realistic materials from multi-view images is a fundamental and challenging task in computer graphics and computer vision. However, due to their highly ill-posed nature, the existing approaches have difficulty decoupling the geometries and materials. Starting from Neural Radiance Field (NeRF)~\cite{mildenhall2021nerf}, the neural rendering-based approaches have brought a significant opportunity for this task and exhibit impressive reconstruction capability. 

These neural reconstruction methods reconstruct the geometries and materials by representing geometries with an implicit function (e.g., density or SDF field) encoded by multi-layer perception (MLP) and modeling the color distribution with a radiance field or a reflectance field. Then, they learn these representations via differentiable rendering, supervised by the multi-view images. These two representations are the key to reconstruction. For the geometry representation, the density-field based methods~\cite{boss2021nerd, zhang2021nerfactor, srinivasan2021nerv, verbin2022ref, jin2023tensoir} have achieved remarkable performance on the novel view synthesis (NVS) and material estimation, while the quality of the reconstructed surface is inferior. The underlying reason is the lack of surface constraints, leading to noisy surfaces and inaccurate surface normals, especially for specular surfaces. In contrast, the SDF-based approaches show superior quality on the reconstructed surfaces. 

Among the previous SDF-based methods, several approaches~\cite{wang2021neus, yariv2021volsdf, zhang2021physg, zhang2022mii, wang2023neus2} rely on radiance-field for surface reconstruction. These methods are effective on non-specular surfaces but fail on objects with strong reflections. The other line of work (e.g., NeRO~\cite{liu2023nero} and NeILF++~\cite{zhang2023neilf++}) model the reflection of surface explicitly with the rendering equation~\cite{kajiya1986rendering}, leading to a much higher reconstruction quality for highly-specular objects. Despite the benefits of reconstructing specular objects, leveraging the reflectance field is not stable enough to optimize and easily falls into local optimum, leading to erroneous geometry. Therefore, no existing approaches can achieve compelling results on objects with arbitrary reflective materials (no translucent). 

In this paper, we propose a novel framework for robust geometry and material reconstruction on top of NeRO~\cite{liu2023nero}, including a geometry reconstruction step, followed by a material estimation step. At the core of our framework is a deep incorporation of the radiance and reflectance fields for arbitrary surface reconstruction. To this end, we propose a roughness-aware combination of these two fields. Furthermore, we introduce a tensorial representation for SDF (i.e., \emph{TensoSDF}), enabling a more detailed reconstructed surface and reducing the training time to about 50\%. Finally, with the reconstructed geometry, we use an explicit-implicit fusion strategy to estimate the material, which uses the explicit mesh as a proxy for efficient ray-intersection computation and the implicit SDF field for more accurate geometric representation, further improving estimated material quality. Consequently, our method can achieve more robust geometry reconstruction and outperform the previous works in terms of relighting quality. To summarize, our main contributions include:
\begin{itemize}
    \item we propose roughness-aware incorporation of the radiance and reflectance field for robust geometry reconstruction on any reflective objects.
    \item we introduce a novel representation -- TensoSDF, that combines the tensorial representation with SDF, enabling more detailed geometry reconstruction and reducing training time. 
    \item we design an explicit-implicit fusion strategy for material estimation, combining the explicit mesh and implicit SDF field, further improving the quality of reconstructed materials.
\end{itemize}

%% file: sec_2-related.tex
\section{Related Work}

\paragraph{Neural geometry reconstruction.}
Existing geometry reconstruction methods from multi-view images include the traditional ones~\cite{campbell2008using, 4270271, strecha2006combined, barron2016fast, bleyer2011patchmatch, gallup2007real, hosni2012fast, richardt2010real, schonberger2016structure} and recent neural approaches~\cite{wang2021neus, yariv2020idr, niemeyer2020dvr, wu2022voxurf, oechsle2021unisurf, li2023neuralangelo, yariv2023bakedsdf}. With the advances in neural rendering~\cite{mildenhall2021nerf}, the neural approaches have shown impressive reconstruction results. In this paper, we focus on neural geometry reconstruction methods. These methods usually use a neural implicit function to express scene geometry, such as the density field~\cite{mildenhall2021nerf, boss2021nerd, zhang2021nerfactor, jin2023tensoir, verbin2022ref}, signed distance field~\cite{yariv2020idr, wang2021neus, wang2023neus2, yariv2023bakedsdf, li2023neuralangelo} or occupancy field~\cite{niemeyer2020dvr, oechsle2021unisurf}, and use a neural color function for differentiable rendering. We categorize these methods into two groups according to their color function: \emph{radiance field} that implicitly encodes the color with the view direction and surface geometry, and \emph{reflectance field} that explicitly incorporates the formulation of the rendering equation~\cite{kajiya1986rendering}.

Among the radiance-field methods, DVR~\cite{niemeyer2020dvr} and IDR~\cite{yariv2020idr} first introduce surface rendering with occupancy function or SDF for 3D geometry reconstruction, respectively. However, these methods need pixel-accurate object masks for surface estimation. Later, NeuS~\cite{wang2021neus}, VolSDF~\cite{yariv2021volsdf} and UNISURF~\cite{oechsle2021unisurf} solve this problem by incorporating the volume-rendering framework from NeRF~\cite{mildenhall2021nerf}, and can obtain decent reconstructed results. With the advent of the explicit representation~\cite{sun2022dvgo, muller2022instant, chen2022tensorf}, Voxurf~\cite{wu2022voxurf}, NeuS2~\cite{wang2023neus2}, Neuralangelo~\cite{li2023neuralangelo} and BakedSDF~\cite{yariv2023bakedsdf} are proposed to improve the geometry quality further and reduce the training time. All these methods have shown remarkable results in reconstructing non-specular objects but fail on objects with strong reflections.

The reflectance-field methods usually require many assumptions to decompose the color into the light and materials. Ref-NeRF~\cite{verbin2022ref} decomposes the color into diffuse and specular terms, and uses integrated direction encoding to improve the NVS quality on reflective objects. TensoIR~\cite{jin2023tensoir} adopts the efficient TensoRF~\cite{chen2022tensorf} framework to explicitly model indirect lights, achieving good geometry reconstruction on diffuse surfaces. These two methods are all based on density field, which has inferior quality on the geometry reconstruction of specular objects. Recently, NeRO~\cite{liu2023nero} designed a novel light representation based on split-sum approximation~\cite{karis2013real}, leading to high-quality geometric results on highly-specular objects. However, NeRO easily falls into the local optimum and causes erroneous geometry. NeILF++\cite{zhang2023neilf++} proposes to marry an incident light field and an outgoing radiance field via physical-based rendering, enables handling specular surfaces and inter-reflections, but often causes over-smooth results and costs a long time to train.

Neither of the above two types of methods can achieve compelling results on objects with arbitrary reflective materials. In contrast, our method is robust and can handle any reflective objects by learning the radiance and reflectance fields together.

\paragraph{Neural material estimation.}
Existing material estimation methods from multi-view images are mainly based on the inverse rendering framework. Most of them~\cite{boss2021nerd, boss2021neural, zhang2021physg} only consider the direct light for rendering to save computation, while lowering the quality. NeRV~\cite{srinivasan2021nerv} considers the visibility and indirect light but requires multiple known lighting conditions. NeRFactor~\cite{zhang2021nerfactor} can handle unknown illumination but without considering indirect light. To model the indirect light, MII~\cite{zhang2022mii} uses Spherical Gaussian to represent the indirect illumination, NDRMC~\cite{hasselgren2022shape} explicitly samples secondary ray with a denoiser, NeILF~\cite{yao2022neilf} and NeILF++~\cite{zhang2023neilf++} represent scene lightings with a neural incident light field, and TensoIR~\cite{jin2023tensoir} resorts to efficient TensoRF~\cite{chen2022tensorf} framework. All these methods improve the quality of estimated materials but produce inferior results on highly reflective objects. NeRO~\cite{liu2023nero} estimates the material using Monte Carlo integration with importance sampling on the fixed mesh, which can produce decent results on reflective objects. However, explicit mesh suffers from geometry degradation, causing apparent biases in the estimated materials. In comparison, we unify the explicit mesh and implicit SDF field for material estimation, further improving the quality of reconstructed materials.

%% file: sec_3-method.tex
% \mycfigure{geo}{geo_pipeline4.pdf}{The network architecture in our geometry reconstruction step. The core of our network is the TensoSDF, consisting of a tensorial encoder and an MLP decoder, which maps the position of sampled points to SDF values and appearance features. This TensoSDF representation is learned by incorporating the radiance and reflectance fields with the \emph{roughness} as a balancing weight. Specifically, the joint loss $l_c$ is designed to combine two color losses $l_\mathrm{rad}$ and $l_\mathrm{ref}$, which are computed by rendering the radiance and reflectance fields, respectively.}

\mycfigure{geo}{geo_pipeline6.pdf}{The network architecture in our geometry reconstruction step. The core of our network is the TensoSDF (on the left), consisting of a tensorial encoder and an MLP decoder, which maps the position of sampled points to SDF values and appearance features. This TensoSDF representation is learned by incorporating the radiance and reflectance fields with the \emph{roughness} as a balancing weight. Specifically, the joint loss $l_c$ is designed to combine two color losses $l_\mathrm{rad}$ and $l_\mathrm{ref}$, which are computed by rendering the radiance and reflectance fields, respectively. The structure of the reflectance field is shown on the right.}

\section{Background}
In this section, we briefly review NeuS and NeRO~\cite{liu2023nero}, since we are related to their work. They both represent the geometry with a neural SDF field, and the underlying surface is the zero-level set of the SDF. Then, they use different ways to reconstruct their surface.

\paragraph{NeuS}
The geometry is reconstructed by performing the volume rendering~\cite{mildenhall2021nerf} with the radiance field. Specifically, given the camera origin $o$ and view direction $d$ of each pixel, NeuS samples $n$ points along the ray $\left \{{p_i=o+t_id \mid t_i > 0, t_{j-1} < t_j}\right \}$, and then aggregates the contribution of these points to form the pixel color $C$:
\begin{equation}
C=\sum_{i=0}^{n}w_ic_i,
\label{eq:volume_rendering}
\end{equation}
where $w_i$ is the weight of the $i$-th point, derived from the SDF value, and $c_i$ is the radiance at this point from the radiance field encoded with a color network. Then, the geometry and the radiance networks can be learned by minimizing the loss between the rendered color $C$ and the ground truth color $C_\mathrm{gt}$:
\begin{equation}
\label{eq:color_loss}
    \ell=\left \| C-C_\mathrm{gt} \right \|^2_2.
\end{equation}

\paragraph{NeRO}
Unlike NeuS, NeRO~\cite{liu2023nero} explicitly incorporates the rendering equation~\cite{kajiya1986rendering} in neural reconstruction using a reflectance field to reconstruct specular objects better. In particular, NeRO reconstructs the geometry by employing a split-sum approximation~\cite{karis2013real} to evaluate the rendering equation. Then, they extract the mesh and perform the Monte Carlo integration to estimate the materials on the extracted meshes. 

Despite NeRO's impressive capability on high specular objects, their model easily falls into the local optimum, resulting in erroneous geometry, due to the extensive assumption from the shading model and light transport. Moreover, the pure MLP-based representation leads to over-smooth geometry and a long training time. %Moreover, the explicit mesh extracted from the SDF field has lost geometric precision, resulting in the quality loss of the estimated material.

\section{Method}
%------------------------%

In this section, we present our approach, following the similar pipeline (a geometry reconstruction step and a material estimation step) as NeRO, but with several key differences, as shown in Fig.~\ref{fig:geo}. We introduce a tensorial representation for SDF (Sec.~\ref{sec:tensor}) and then propose to incorporate the radiance and reflectance fields for surface reconstruction (Sec.~\ref{sec:rough-aware}). After reconstructing the geometry, we unify the explicit mesh and the implicit SDF field for material estimation (Sec.~\ref{sec:mesh-sdf}). 
% Note that the incorporation of radiance and reflectance fields is a general idea, and we present it first.

% \myfigures{roughness}{rough-aware}{
% \added{check later.} Results of the radiance field, reflectance field and roughness in geometry reconstruction stage. The radiance field can model the low-frequency angular effects, while the reflectance field is better to express the high-frequency reflections. The roughness is used to balance the learning of the two fields.}{0.75\linewidth}

% \myfigures{roughness}{rough-aware}{\revised{The heat map of roughness. The roughness indicates the reflection on the surfaces and thus adaptively adjusts the learning weights of the radiance and reflectance fields.}}{0.75\linewidth}

\subsection{TensoSDF representation}
\label{sec:tensor}
% To represent the geometry surface of the object, our goal is to find an representation that maps any 3D position $p$ in the scene to its SDF value $s$, while this representation needs to be high-capacity and efficient. Pure MLP-based representation is hard to meet the above two requirements. Thanks to recent explicit representations such as volume grid~\cite{sun2022dvgo}, hash grid~\cite{muller2022instant} and tensor grid~\cite{chen2022tensorf}, which can achieve this goal. Considering storage overhead and versatility, we introduce compact tensorial representation into our architecture. 

% Thanks to the roughness-aware radiance and reflectance field, we are able to reconstruct any reflective object. However, we notice that the reconstructed objects are over-smoothing, and the training time is exhausting. Hence, we introduce a new representation for the SDF -- TensoSDF, combining the tensorial representation with SDF to replace the original MLP encoded ones. A well-known issue of the tensorial representation is the noisy reconstruction surface, due to the lack of global correlation. To alleviate this problem, we further introduce two smoothness priors to reduce the noise.

To represent the geometry of the object, our goal is to find an high-capacity and efficient representation. Hence, we introduce a new representation for the SDF -- TensoSDF, combining the tensorial representation with SDF to replace the original MLP encoded ones. A well-known issue of the tensorial representation is the noisy reconstruction surface, due to the lack of global correlation. To alleviate this problem, we further introduce two smoothness priors to reduce the noise.

%To represent the geometry of the object, our goal is to find an high-capacity and efficient representation that maps any 3D position $p$ in the scene to its SDF value $s$. To this end, We propose TensoSDF, combing the tensorial representation with SDF, enabling a more detailed reconstructed geometry. However, due to the lack of global correlation, the explicit representation is easy to cause noisy reconstructed surface. To alleviate this problem, we use two smoothness priors to reduce the noise.

\paragraph{Tensorial representation.}
We use the Vector-Matrix factorization proposed by TensoRF~\cite{chen2022tensorf} as our tensorial encoder. Since SDF is a continuous function, directly using pure explicit representation to model the SDF field leads to unstable training. To address this issue, we use a small MLP as a decoder after tensorial encoding. Moreover, different from TensoRF, which uses two separate tensor grids to encode the geometry and appearance, we leverage one shared tensorial encoder and MLP decoder, which can enhance the correlation between the geometry and appearance. Our tensorial representation is formulated as follows:

\begin{align}
\label{eq:tensor}
    V_p & = \mathrm{v}^X_k \circ \mathrm{M}^{YZ}_k \oplus \mathrm{v}^Y_k \circ \mathrm{M}^{XZ}_k \oplus \mathrm{v}^Z_k \circ \mathrm{M}^{XY}_k, \\
    \{s,v_f\} & = \Theta(V_p, p),
\end{align}

where $\mathrm{v}^m_k$ and $\mathrm{M}^{\widetilde m}_k$ represent the $k$-th vector and matrix factors of their corresponding spatial axes $m$, and $\widetilde m$ denotes the two axes orthogonal to $m$ (e.g., $\widetilde X=YZ$). $\circ$ and $\oplus$ represent the element-wise multiplication and concatenation operations. $V_p$ is the latent vector from the tensorial encoder and then is decoded with the position $p$ by a tiny MLP $\Theta$ to get the SDF value $s$ and the appearance feature $v_f$.

\paragraph{Smoothness priors.}
With the TensoSDF, we further introduce two strategies during training and inference to alleviate the noise issue. First, during training, we introduce a Gaussian smooth loss on the tensor grid, which is expressed as follows:
\begin{equation}
\label{eq:lg}
    \ell_g=\sum_{k=1}^{K} \left \| G(\mathrm{M}_k \mid k_g, \sigma_g)-\mathrm{M}_k \right \|^2_2 + \left \| G(\mathrm{v}_k \mid k_g, \sigma_g)-\mathrm{v}_k \right \|^2_2,
\end{equation}
where $G$ is denoted as a 2D Gaussian convolution with kernel size $k_g$ and standard deviation $\sigma_g$. In this way, the local consistency of the tensor grid can be enhanced.

After training the SDF field, we construct a two-layer mipmap of the tensor grid by performing the bilinear interpolation. When extracting the mesh by Marching Cube~\cite{lorensen1998marching}, we compute the final SDF $\hat{s}$ by blending the SDF value $s$ by Eqn.~(\ref{eq:tensor}) from the base tensor grid and the other one $s'$ from the top layer with weight $\alpha$: 
\begin{align}
    \hat{s}&=(1-\alpha )\cdot s+\alpha \cdot s',\\
        s'&= \Theta(V'_p, p),
\end{align}
where $V'_p$ is the latent vector at the top layer. As a result, the extracted mesh is much smoother while keeping the details.
 % Then, it is used for material estimation in the second stage.
%Second, when inferring the trained SDF field to extract the mesh by Marching Cube~\cite{lorensen1998marching}, we blend two SDF values as the final result. One is evaluated by the Eq.~\ref{eq:tensor} from the base tensor grid. For the other, we use bilinear interpolation to construct one-layer mipmap of the base tensor grid. Then, we calculate its latent vector $V'_p$ on this layer to feed it into the MLP decoder to get the second SDF value $s'$. It can be formulated as:
%\begin{align}
%    \hat{s}&=(1-\alpha )\cdot s+\alpha \cdot s'\\
%        s'&= \Theta(V'_p, p)
%\end{align}
%where $\hat{s}$ is the final SDF value for extracting the mesh and $\alpha$ is the blending coefficient. 

\subsection{Incorporating the radiance and reflectance fields}
\label{sec:rough-aware}

The explicit incorporation of the rendering equation in NeRO~\cite{liu2023nero} enables a high-quality reconstruction of specular objects since it can express the high-frequency angular effects better than using the implicit color function directly (i.e., radiance field). However, we notice that this reflectance field is unstable and falls into the local optimum, resulting in erroneous geometry. The underlying reason is that for objects with smooth angular effects, the implicit color function better fits the multi-view observations due to the extensive assumptions of the shading model and light transport for the explicit rendering. Hence, we propose incorporating both the reflectance and radiance fields. The main question is how to fuse these two fields. One straightforward way is weighting them with a fixed weight; however, it shows inferior reconstructed quality. To this end, we propose a simple yet effective way, leveraging the roughness at each point as a balance factor.

%To design a robust method which can reconstruct geometry of any reflective objects, we propose to learn the radiance field and the reflectance field together. The radiance field correlates the color with the view direction and surface geometry without any assumption on the light and shading model, can better represent the low-frequency angular effects. And the reflectance field explicitly incorporate the formulation of the rendering equation~\cite{kajiya1986rendering}, can model the high-frequency specular effects, as shown in Fig.~\ref{fig:roughness}. We starts to present the radiance field and the reflectance field used in our pipeline, and then introduce our roughness-aware learning strategy for balancing the two fields.

\paragraph{Radiance field \& reflectance field.}
We follow the NeuS~\cite{wang2021neus} to model our radiance field:
\begin{equation}
\label{eq:rad_net}
    c_{\mathrm{rad}} = \Theta_{\mathrm{rad}}(p, n, v_f, d).
\end{equation}
Here, $\Theta_{rad}$ is the radiance MLP, $p$ is a sample position, $n$ is the normal at $p$ derived from the SDF value, $v_f$ is the appearance features output from the TensoSDF (Eqn.~(\ref{eq:tensor})) and $d$ is the view direction.

Then, we utilize the reflectance field from NeRO~\cite{liu2023nero}, as shown on the right in Fig.~\ref{fig:geo}, which can be formulated as:
\begin{align}
\label{eq:reflectance}
\{a, m, r\} & = \Theta_{\mathrm{mat}}(p, v_f),\\
L_\mathrm{d} & = \Theta_{\mathrm{direct}}(n),\\   
L_\mathrm{ind} & = \Theta_{\mathrm{indirect}}(p,w_r, r), \\
u & = \Theta_{\mathrm{occ}}(p, w_r),\\
c_{\mathrm{ref}} & = \rho_{\mathrm{diff}}\cdot L_\mathrm{d} + \rho_{\mathrm{spec}}\cdot ((1-u)\cdot L_\mathrm{d}+u\cdot L_\mathrm{ind}),
\end{align}
where $\Theta_{\mathrm{mat}}$ is the material MLP which outputs albedo $a$, metallic $m$ and roughness $r$. $\Theta_{\mathrm{direct}}$ and $\Theta_{\mathrm{indirect}}$ are the direct-light MLP and indirect-light MLP, respectively. $u$ is the occlusion probability from the occlusion MLP $\Theta_{\mathrm{occ}}$. $w_r$ is the reflected view direction around the normal $n$. $\rho_{\mathrm{diff}}$ and $\rho_{\mathrm{spec}}$ are the diffuse and specular terms from the microfacet BRDF~\cite{cook1982reflectance} respectively.

\paragraph{Roughness-aware learning.}
With both the radiance color $c_{\mathrm{rad}}$ and reflectance color $c_{\mathrm{ref}}$ at each sampled point, we compute two aggregated pixel colors by Eqn.~(\ref{eq:volume_rendering}): $C_{\mathrm{rad}}$ and $C_{\mathrm{ref}}$. Then, we compute their own loss w.r.t. the multi-view images by Eqn.~(\ref{eq:color_loss}): $\ell_\mathrm{rad}$ and $\ell_\mathrm{ref}$, respectively. To balance the learning of the radiance field and reflectance field, we introduce roughness from Eqn.~(\ref{eq:reflectance}) as the balancing factor into the final loss, formulated as:
\begin{equation}
\label{eq:lc}
    \ell_\mathrm{c} = r\cdot \ell_{\mathrm{rad}} + (1-r) \cdot \ell_{\mathrm{ref}}.
\end{equation}

The gradient of roughness $r$ is detached here for stable optimization. In this way, our model can adaptively adjust the learning weights of two fields according to the surface roughness, enabling robust reconstruction of any reflective objects.

\subsection{Unifying mesh and SDF for material estimation}
\label{sec:mesh-sdf}

We now have reconstructed the faithful geometry but with a rough material. A simple way to achieve more accurate material is following NeRO directly by evaluating the Monte Carlo integration with importance sampling using the extracted mesh due to its ray-intersection efficiency. However, using the extracted mesh for material estimation leads to geometry degradation. On the other hand, using the implicit SDF field with volume rendering is too time-consuming, even under the tensorial representation, due to the exponential-increasing computation and memory cost for the indirect lighting. 

%In this stage, our goal is to reconstruct more accurate material results. Following NeRO~\cite{liu2023nero}, we apply Monte Carlo integration with importance sampling to reconstruct the materials. However, NeRO evaluates the rendering equation only on the fixed mesh extracted from the SDF field, which sacrifices the geometry accuracy for time and memory. If we directly use the implicit SDF field with volume rendering, even under the tensorial representation, we can not afford the exponential-increasing computation and memory when considering indirect light. 

\myfigures{proxy}{mesh_sdf2}{Illustration of our mesh-SDF fusion strategy. First, we perform ray-mesh intersection to get a rough hit point. Then, we sample $m$ points within a fixed distance inside and outside the surface.}{0.8\linewidth}

To avoid this problem, we use explicit mesh as a proxy to get a rough hit point by fast ray-mesh intersection and then perform volume rendering on the neighbor of the hit surface, as shown in Fig.~\ref{fig:proxy}. Here, we sample $m$ points within the distance of $4u$ inside and outside the surface, where $u$ is the unit size of the tensor grid. Then, we perform volume rendering to get the accurate hit point $\hat{p}$ and its surface normal $\hat{n}$:
\begin{align}
\hat{t}&=\sum_{i=0}^{m}w_it_i, \\ 
\hat{n}&=\sum_{i=0}^{m}w_in_i, \\
\hat{p}&=o+\hat{t} d.
\end{align}
Here $t_i$ and $n_i$ are the depth and normal of $i$-th sampled point.

Thanks to the mesh and implicit SDF combination, our method can achieve more accurate material estimation while maintaining time efficiency.

%% file: sec_4-implementation.tex
\section{Implementation details}

% In this section, we present our datasets and training details here and more details can be found in the supplementary. We will release our codes and datasets.

In this section, we present our datasets, network structures, and training details.

\paragraph{Datasets.}
 To validate our method, we propose a new synthetic dataset consisting of six scenes with various typical material types (diffuse, glossy, and specular). We render these scenes to generate their ground-truth images, normal maps, and relighting images under five environment lights using the Cycles renderer in Blender. Moreover, we also evaluate our method on datasets from TensoIR~\cite{jin2023tensoir} and NeRO~\cite{liu2023nero}, which mainly focus on diffuse or specular scenarios, respectively. For the real datasets, we use the NeILF-HDR dataset from NeILF++~\cite{zhang2023neilf++} and Stanford-ORB dataset~\cite{kuang2024stanford}. We adopt the ACES tone mapping~\cite{gatta2002ace} to obtain the low-dynamic range (LDR) images from the NeILF-HDR dataset, since the high-dynamic range (HDR) images are unnecessary in our experiments. The Stanford-ORB dataset includes ground-truth captured meshes and relighting images, and we choose its LDR images for our experiments.

%\added{Note that structure first, then loss, then training.}

\paragraph{Network structures.}

The TensoSDF has a resolution of $512\times 512$ with the feature channels set as 36. The decoder is a two-layer MLP with a width of 128. The MLP networks in the radiance and reflectance fields (Eqn.~(\ref{eq:rad_net}) and Eqn.~(\ref{eq:reflectance})) all have three layers with widths of 128. 

\paragraph{Training.}
We define different losses for the geometry reconstruction and the material estimation stages. The loss function at the geometry reconstruction stage includes a color loss $\ell_c$ (Eqn.~(\ref{eq:lc})), a Gaussian smooth loss $\ell_g$ (Eqn.~(\ref{eq:lg})), and an Eikonal loss $\ell_e$ for valid SDF learning~\cite{gropp2020eikonal}. We also use a total variation (TV) loss $\ell_{t}$ from TensoRF~\cite{chen2022tensorf}, occlusion loss $\ell_{o}$ and stabilization loss $\ell_{s}$ from NeRO~\cite{liu2023nero} for the reflectance field optimization. Furthermore, depending on whether the object mask is available, we use a mask loss for better convergence or a Hessian loss~\cite{zhang2022critical} $\ell_h$. The final loss is:
\begin{equation}
    \ell = \ell_c + \lambda_e\ell_e + \lambda_g\ell_g + \lambda_h\ell_h + \lambda_t\ell_t + \ell_o + \ell_s,
\end{equation}
where $\lambda$ is the corresponding weight of each loss (we set $\lambda_e=\lambda_t=0.1$, $\lambda_g=1e^{-5}$ and $\lambda_h=0.5$ or $5e^{-4}$ for mask loss or Hessian loss in practice).

The loss function at the material estimation stage includes a color loss $\ell_c$ and a material regularization loss $\ell_m$ from NeRO~\cite{liu2023nero}:
\begin{equation}
    \ell = \ell_c + \lambda_m\ell_m,
\end{equation}
where $\lambda_m$ is the loss weight of the regularization (0.1 in practice).

During the training of the geometry reconstruction stage, we use a coarse-to-fine training strategy by setting the initial grid resolution as $128\times 128$ and increasing to the final resolution, i.e., $512\times 512$. At the first 20k steps, we only learn the reflectance field to get a rough shape and an initial roughness. Then, we compute two pixel colors from the radiance and reflectance fields. We use these two colors to compute two separate losses and combine the two losses with the roughness-aware strategy by Eqn.~(\ref{eq:color_loss}). During inference, we use the reflectance field or the recovered materials to compute the final color.

We use Adam optimizer~\cite{kingma2014adam} in PyTorch~\cite{paszke2019pytorch} with initial learning rates of 0.01 for the tensor grids and 0.001 for all the MLP networks, and use cosine learning rate scheduler with the target decay ratio 0.05. The training epochs for the geometry stage are 180k, and for the material stage are 100k, which take an average of 4 hours and 1.5 hours on a single RTX 4090 GPU, respectively.

%% file: sec_5-results.tex
\section{Results}
\label{sec:results}

In this section, we first evaluate our geometry reconstruction (Sec.~\ref{sec:eval_geo}) and material estimation (Sec.~\ref{sec:eval_mat}) on the synthetic datasets. Then, we present reconstruction results on the real datasets (Sec~\ref{sec:real}). We perform ablation studies to verify the effectiveness of our design (Sec.~\ref{sec:ab}). Finally, we give the discussion of our limitations (Sec.~\ref{sec:limit}). More results can be found in our supplementary materials and video.

\subsection{Reconstructed geometry evaluation}
\label{sec:eval_geo}

To evaluate the geometry reconstruction quality, we adopt the mean angular error (MAE) between the reconstructed normal and the ground-truth normal as the metric and use the Chamfer distance (CD) metric on the NeRO~\cite{liu2023nero} dataset. We compare our method with NeuS~\cite{wang2021neus}, TensoIR~\cite{jin2023tensoir}, NeRO~\cite{liu2023nero} and NeILF++~\cite{zhang2023neilf++}. For all these methods, we use their official implementations. 

In Fig.~\ref{fig:geo_syn_comp}, we compare our method with the other four methods on our synthetic dataset. By comparison, our method can not only reconstruct the surfaces with strong reflection in all three scenes but also recover the detailed geometries (the grid in the \textsc{Compressor} scene and the tube in the \textsc{Helmet} scene). TensoIR~\cite{jin2023tensoir} has difficulties in handling reflective surfaces, since it represents the geometry with the density field. NeRO~\cite{liu2023nero} shows impressive reconstruction on simple specular objects but easily falls into local optimum for complex scenes, leading to incorrect geometries. NeuS~\cite{wang2021neus} can reconstruct most non-specular surfaces except the ones with strong reflections. NeILF++~\cite{zhang2023neilf++} can reconstruct geometries reasonably on various materials but loses geometry details and needs a long training time (more than 12 hours for each scene). We also report the quality metrics on our synthetic dataset and the average training time in Tab.~\ref{tab:geo_res}. Our method outperforms the other four methods on the average MAE metrics across six scenes, but has slightly higher errors than NeuS on the \textsc{Dragon} and \textsc{Motor} scenes, mainly due to the noise problem. Thanks to the tensorial representation, our method achieves shorter training times than the other three methods and equals TensoIR~\cite{jin2023tensoir}.

We also conduct the comparison on the datasets from NeRO~\cite{liu2023nero} and TensoIR~\cite{jin2023tensoir}, as shown in Fig.~\ref{fig:nero_syn_geo_comp} and Fig.~\ref{fig:tensoir_syn_geo_comp}. Our method can handle both diffuse and specular scenes, demonstrating its robustness. Meanwhile, our method exhibits the most detailed geometry details, thanks to the TensoSDF representation.

\mycfigure{geo_syn_comp}{geo_syn_comp.pdf}{Comparison of geometry reconstruction among our method, TensoIR~\cite{jin2023tensoir}, NeRO~\cite{liu2023nero}, NeILF++~\cite{zhang2023neilf++} and NeuS~\cite{wang2021neus} on our synthetic dataset.}

\mycfigure{nero_syn_geo_comp}{nero_dataset_comp_geo.pdf}{Comparison of geometry reconstruction among our method, NeILF++~\cite{zhang2023neilf++}, NeRO~\cite{liu2023nero} and NeuS~\cite{wang2021neus} on the scene from NeRO~\cite{liu2023nero}. There is no available ground-truth mesh in NeRO dataset, and the CD$\downarrow$ loss is computed on the point cloud following NeRO~\cite{liu2023nero}.}

\subsection{Estimated material evaluation}
% \subsection{Estimated material validation}
\label{sec:eval_mat}

Since different methods use different shading models, comparing the reconstructed materials directly is difficult. Following NeRO~\cite{liu2023nero}, we relight the estimated materials and compare their results with the ground truth as one effective way for validation. Specifically, we first extract the recovered meshes and materials, and re-illuminate the scene with five new environment lights in, each rendering 20 evenly distributed relighted images. We use peak signal-to-noise ratio (PSNR), structural similarity index (SSIM)~\cite{wang2004ssim}, and learned perceptual image patch similarity (LPIPS)~\cite{zhang2018lpips} metrics to measure the quality of results. We compare our method against three others: TensoIR~\cite{jin2023tensoir}, NeRO~\cite{liu2023nero} and NeILF++~\cite{zhang2023neilf++} on our synthetic dataset. We scale the relighted images with a global scalar for all the methods, as done in TensoIR~\cite{jin2023tensoir}.

The quantitative measurements are shown in Tab.~\ref{tab:relight_res} for PSNR and SSIM, and in our supplementary for LPIPS. We also provide the visual results in Fig.~\ref{fig:relight_syn_comp}. By comparison, our method achieves more plausible relighting results than other methods. NeRO~\cite{liu2023nero} mainly suffers from inaccurate reconstructed geometry, resulting in inferior relighting quality. On the other hand, NeILF++~\cite{zhang2023neilf++} and TensoIR~\cite{jin2023tensoir} produce less-specularity materials and struggle to keep the high-frequency effects, due to their Spherical Gaussian approximation on the BRDF or environment light.

We visualize the estimated materials (albedo, roughness and metallic) on \textsc{Helmet} and \textsc{Robot} scenes, as shown in Fig.~\ref{fig:mat_decomp}. We scale the albedo results by a global scalar, as done in TensoIR~\cite{jin2023tensoir}. By comparison, our method can achieve the closest estimated albedo maps, and reasonable metallic and roughness maps, leading to credible relighting results.

\begin{table}
  \centering
  \small
  \setlength\tabcolsep{2pt}
  \caption{Geometry reconstruction quality in terms of normal MAE$\downarrow$ on our synthetic dataset. \textbf{Bold} means the best performance and \underline{underline} means the second best.}
  \label{tab:geo_res}
  \begin{tabular}{cccccc}
    \hline
            & NeuS & TensoIR & NeRO & NeILF++ & Ours \\
    \hline
    Rover          & \underline{3.25} & 3.40 & 5.31 & 3.52 & \textbf{3.21}  \\
    Dragon         & \textbf{2.33} & \underline{2.50} & 3.99 & 3.26 & 2.59  \\
    Motor          & \textbf{3.59} & 3.78 & 4.78 & 4.54 & \underline{3.70}  \\
    Helmet         & 3.45 & 3.45 & 8.27 & \underline{3.14} & \textbf{2.74}   \\
    Robot          & 2.65 & 2.73 & 5.57 & \underline{2.59} & \textbf{2.04}    \\
    Compressor     & \underline{5.09} &  5.37 & 9.18 & 6.06 & \textbf{3.48}    \\
\hline
    Avg. MAE    & \underline{3.39} & 3.54 & 6.18 & 3.85 & \textbf{2.96}    \\
    Avg. training time  & \underline{6 hrs} & \textbf{4 hrs} & 8 hrs & 12 hrs & \textbf{4 hrs}    \\
\hline
\end{tabular}
\end{table}

\begin{table}
  \centering
  \small
  \setlength\tabcolsep{2pt}
  \caption{Relighting quality in terms of PSNR and SSIM on our synthetic dataset. \textbf{Bold} means the best performance and \underline{underline} means the second best.}
  \label{tab:relight_res}
  \begin{tabular}{ccccc}
    \hline
            & \tabincell{c}{TensoIR\\PSNR / SSIM}  & \tabincell{c}{NeRO\\PSNR / SSIM} & \tabincell{c}{NeILF++\\PSNR / SSIM} & \tabincell{c}{Ours\\PSNR / SSIM} \\
    \hline
    Rover          & 24.000 / \underline{0.918} & \underline{24.015} / 0.914 & 23.774 / 0.911 & \textbf{26.754 / 0.935}  \\
    Dragon         & 25.104 / 0.895 & \underline{25.644} / \underline{0.919} & 24.099 / 0.901 & \textbf{27.899 / 0.936}  \\
    Motor          & 19.219 / 0.906 & \underline{22.158} / \underline{0.917} & 20.142 / 0.894 & \textbf{22.754 / 0.930}  \\
    Helmet         & \underline{25.140} / 0.901 & 22.587 / 0.881 & 24.001 / \underline{0.906} & \textbf{28.126 / 0.934}   \\
    % Buddha         & \underline{24.456} / \underline{0.929} & 19.453 / 0.905 & 23.453 / 0.923 & \textbf{27.240 / 0.938}   \\
    Robot          & \underline{26.031} / \underline{0.928} & 23.194 / 0.913 & 22.696 / 0.915 & \textbf{26.242 / 0.940}    \\
    Compressor     & 20.753 / 0.868 & \underline{21.624} / \underline{0.878} & 19.740 / 0.844 & \textbf{24.049 / 0.916}    \\
\hline
    Avg.           & \underline{23.375} / 0.903 &  23.204 / \underline{0.904} & 22.41 / 0.895 & \textbf{25.971 / 0.932}   \\
\hline

\end{tabular}
\end{table}

\subsection{Results on real data}
\label{sec:real}
We first evaluate our method on the real dataset from NeILF++~\cite{zhang2023neilf++}, as shown in Fig.~\ref{fig:real_comp}. We compare our method with NeRO~\cite{liu2023nero} and NeILF++~\cite{zhang2023neilf++} on geometry reconstruction. Note that the input images of our method and NeRO~\cite{liu2023nero} are LDR images, while NeILF++~\cite{zhang2023neilf++} uses the HDR images following the official implementation. By comparison, our method is able to handle surfaces with strong highlights, while NeILF++~\cite{zhang2023neilf++} fails (pointed by the red arrow in \textsc{Qilin} and \textsc{Luckycat} scenes). NeRO~\cite{liu2023nero} can also produce reasonable results but loses many geometric details, such as the characters in the \textsc{Luckycat} and \textsc{Brassgourd} scenes. Our method produces plausible relighting results (shown in the right column of Fig.~\ref{fig:real_comp}), demonstrating the accuracy of estimated materials.

We also evaluate our method on recent Stanford-ORB~\cite{kuang2024stanford} dataset. We choose five typical objects with masks. For each object, we randomly select one environment light for training and the other two lights for relighting. The quantitative measurements of the geometry reconstruction and relighting results are shown in Tab.~\ref{tab:real_data}. Our method outperforms NeRO~\cite{liu2023nero} on all five scenes. The visual results are shown in Fig~\ref{fig:real_orb}. NeRO~\cite{liu2023nero} produces erroneous surfaces in the \textsc{Teapot} and \textsc{Cactus} scenes, and over-smoothed results in the \textsc{Gnome} scene. Our method can restore decent object geometries and materials, leading to plausible relighting results.

\mycfigure{tensoir_syn_geo_comp}{tensoIR_dataset_comp_geo.pdf}{Comparison of geometry reconstruction among our method, NeILF++~\cite{zhang2023neilf++}, NeRO~\cite{liu2023nero} and NeuS~\cite{wang2021neus} on scenes from TensoIR~\cite{jin2023tensoir}.}

\mycfigure{relight_syn_comp}{relight_syn_comp2.pdf}{Comparison of relighting results among our method, TensoIR~\cite{jin2023tensoir}, NeILF++~\cite{zhang2023neilf++} and NeRO~\cite{liu2023nero} on our synthetic dataset.}

\clearpage

\mycfigure{mat_decomp}{mats_new.pdf}{Comparison of the estimated materials among our method, NeILF++~\cite{zhang2023neilf++} and NeRO~\cite{liu2023nero} on \textsc{Helmet} and \textsc{Robot} scenes. The PSNR metric is calculated between the estimated albedo after re-scaling and ground-truth albedo, as done in TensoIR~\cite{jin2023tensoir}.}

\mycfigure{real_comp}{real_relight_neilfppdata.pdf}{Visual comparison of geometry reconstruction among our method, NeRO~\cite{liu2023nero} and NeILF++~\cite{zhang2023neilf++} and our relighting results on the real data from NeILF++~\cite{zhang2023neilf++}. Note that the input images of our method and NeRO~\cite{liu2023nero} are LDR images, while NeILF++~\cite{zhang2023neilf++} uses the HDR images following the official implementation.}

\clearpage

\mycfigure{real_orb}{real_orb.pdf}{Comparison of the geometry reconstruction (Left) and relighting results (Right) between our method and NeRO~\cite{liu2023nero} on Stanford-ORB~\cite{kuang2024stanford} dataset. The PSNR metric is averaged on two environment lights.}

\mycfigure{geo_ab}{geo_ablation3.pdf}{Ablation studies on the geometry reconstruction. Introducing the TensoSDF (b) can reconstruct more details than NeRO~\cite{liu2023nero} (a) but with problematic surfaces. After combining the radiance field and reflectance field with a fixed balancing weight of 0.5 (c), the surface quality can be improved, but still with some flaws. Finally, using the roughness as the weight (d) achieves the best quality.}

\mycfigure{roughness_ab}{roughness_ab.pdf}{Ablation studies on the different balancing weights. With the radiance filed only (a), i.e., $r=1$, the reconstructed geometry shows apparent flaws on the reflective surface. With the balancing weights decrease (b-d), the reconstructed quality gradually improves. Using the roughness as the weight (e) achieves the best quality.}

\clearpage

\begin{table}
  \centering
  \small
  \setlength\tabcolsep{2pt}
  \caption{Geometry reconstruction quality in terms of CD$\downarrow$ ($\times10^{-4}$) metric and relighting quality in terms of PSNR, SSIM and LPIPS$\downarrow$ metrics on the Stanford-ORB~\cite{kuang2024stanford} dataset. \textbf{Bold} means the best quality.}
  \label{tab:real_data}
  \begin{tabular}{ccccc}
    \hline
            & \tabincell{c}{CD$\downarrow$\\NeRO / Ours}  & \tabincell{c}{PSNR\\NeRO / Ours} & \tabincell{c}{SSIM\\NeRO / Ours} & \tabincell{c}{LPIPS$\downarrow$\\NeRO / Ours} \\
    \hline
    Teapot        & 0.395 / \textbf{0.265}  & 26.369 / \textbf{29.971} & 0.978 / \textbf{0.986} & 0.0344 / \textbf{0.0276}  \\
    Gnome         & 0.892 / \textbf{0.396}  & 26.465 / \textbf{28.488} & 0.929 / \textbf{0.956} & 0.1082 / \textbf{0.0947}  \\
    Cactus        & 4.144 / \textbf{0.199} & 29.081 / \textbf{31.351} & 0.977 / \textbf{0.984} & 0.0399 / \textbf{0.0362}  \\
    Car           & 0.588 / \textbf{0.349}  & 27.763 / \textbf{28.010} & 0.979 / \textbf{0.981} & 0.0350 / \textbf{0.0367}   \\
    Grogu         & 4.804 / \textbf{3.968}   & 27.081 / \textbf{30.595} & 0.983 / \textbf{0.990} & 0.0396 / \textbf{0.0362}    \\
\hline
    Avg.          & 2.165 / \textbf{1.035} & 27.352 / \textbf{29.683}  & 0.969 / \textbf{0.979} & 0.0514 / \textbf{0.0463} \\
\hline

\end{tabular}
\end{table}

\begin{table}
  \small
  \centering
  \caption{The impact of our mesh-SDF fusion strategy for the relighting results on the \textsc{Dragon} and \textsc{Rover} scenes. \textbf{Bold} means the best quality.}
  \label{tab:mats_ab}
  \begin{tabular}{l|c|c|c|c}
    \hline
    Scene  & mesh-SDF fusion  & PSNR & SSIM & LPIPS$\downarrow$  \\
    \hline
    \multirow{2}*{Dragon} & \XSolidBrush    & 27.396 & 0.935 & 0.081 \\
                          & \CheckmarkBold  & \textbf{27.899} & \textbf{0.936} & \textbf{0.078}\\
    \hline
    \multirow{2}*{Rover}  & \XSolidBrush    & 26.466 & 0.933 & 0.060 \\
                          & \CheckmarkBold  & \textbf{26.754} & \textbf{0.935} & \textbf{0.059} \\
  \hline
\end{tabular}
\end{table}

\subsection{Ablation study}

\label{sec:ab}

We validate the key components of our method in Fig.~\ref{fig:geo_ab} step by step. Starting from NeRO~\cite{liu2023nero}, we replace its pure MLP-based geometry representation with our TensoSDF. The TensoSDF can reconstruct more geometric details and significantly improves the quality (1.65 in MAE). Moreover, TensoSDF reduces the training time to about 50\%, from 8 hours to less than 4 hours for each scene in our synthetic dataset. However, the reconstructed geometry still exhibits a noticeable dent on the surface. Then, after introducing both radiance and reflectance fields with a constant weight (0.5 in practice), the surface artifacts are alleviated. Finally, replacing the constant weight with the estimated roughness forms our complete solution and closely matches the ground truth.

To further study the effectiveness of our roughness-aware balancing strategy, we provide ablation studies on different balancing weights, as shown in Fig.~\ref{fig:roughness_ab}. We first set the weight $r=1$, i.e., the radiance field only, showing apparent flaws on the reflective surfaces. With the weights decreasing and approaching the surface roughness, the reconstructed quality becomes better. When setting the balancing weight as the roughness, the quality achieves the best.

Another critical strategy in our material estimation stage is the mesh-SDF fusion. We show its impact on the estimated roughness (Fig.~\ref{fig:mat_ab}) and relighting results (Tab.~\ref{tab:mats_ab}). Without the mesh-SDF fusion, there are many apparent biases, mainly caused by the geometry degradation. The mesh-SDF fusion can reduce these biases, further improving the quality of reconstructed materials and relighting results.

We also present the ablation of our two smoothness priors in Fig.~\ref{fig:smooth_ab}. Directly using the explicit representation without any smoothness priors easily causes noise on the reconstructed geometry, especially on flat surfaces. The mipmap blending strategy of the tensor grid alleviates this problem during inference, and equipping the Gaussian smooth loss in training can produce smoother results.

\myfigures{mat_ab}{mat_ablation4}{The impact of our mesh-SDF fusion strategy for estimated roughness. The result without the mesh-SDF fusion (left column) shows obvious bias, while applying the mesh-SDF fusion (middle column) can better match the ground truth (right column).}{1.0\linewidth}

\myfigures{smooth_ab}{smooth_ab}{The impact of our two smoothness priors. Reconstructing object geometry using explicit representation without any smooth priors causes noise on the smooth surface (a), while applying the mipmap blending strategy alleviates this problem (b). Finally, introducing the Gaussian smooth loss in training (c) produces the smoothest result.}{1.0\linewidth}

\subsection{Discussion and limitations}
\label{sec:limit}
Although the high-capacity TensoSDF improves the geometric details and speeds up the training, it still introduces some limitations. One problem is the trade-off between fine details and high-frequency noise. Explicit representations are more prone to over-fitting locally, which inevitably leads to noise. A common mitigation is to use additional smooth regularization, such as our Gaussian smooth loss and mipmap blending strategy. However, smooth priors also easily cause the loss of details, as shown in Fig.~\ref{fig:limitation}. We leave this problem for future work.

The other problem is the trade-off between the quality and storage. Though the tensorial representation is compact compared to volume-based representations, the storage cost is higher than MLP-based methods. We analyze this trade-off on different resolutions of the tensor grid, as shown in Tab.~\ref{tab:storage}. With the resolution increasing, the reconstructed geometry quality improves while the storage increases. However, even on resolution 200 $\times$ 200, our method can still produce decent results. One possible solution to reduce storage further is pruning; we leave this for future work.

\myfigures{limitation}{limitations2}{The trade-offs between the geometric details and noise. The smooth priors can alleviate the noise problem introduced from the explicit representation (first row) while also causing the loss of details (second row).}{1.0\linewidth}

\begin{table}
  \small
  \centering
  \caption{The trade-offs between the geometry quality and storage. \textit{Ours-200} means that the final resolution of the tensor grid is 200 $\times$ 200. With the resolution increasing, the normal MAE of the reconstructed result decreases, while the model size becomes larger. The results are from the \textsc{compressor} scene. \textbf{Bold} means the best performance and \underline{underline} means the second best.}
  \label{tab:storage}
  \begin{tabular}{l|c|c|c|c|c}
    \hline
                & NeRO & TensoIR & Ours-200 & Ours-300 & Ours-512 \\
    \hline
    MAE$\downarrow$        & 9.18 & 5.09  & 3.96 & \underline{3.64} & \textbf{3.48} \\
    Size (MB)$\downarrow$   & \textbf{8.3} & 70.5 & \underline{17.9} & 37.6 & 109.0 \\ 
  \hline
\end{tabular}
\end{table}

%% file: sec_6-conclusion.tex
\section{Conclusion and Future work}

% \paragraph{Conclusion.}
In this paper, we have presented a new framework for robust geometry and material reconstruction. The geometry is represented by a tensorial-encoding SDF field, which obtains more geometric details and reduces the training time. By incorporating the radiance and reflectance fields in a roughness-aware manner, our method is able to reconstruct any reflective objects robustly. Furthermore, we have designed a mesh-SDF fusion strategy for material estimation, improving the quality of estimated materials. We have demonstrated that our method can achieve more robust geometry reconstruction and outperform the existing state-of-the-art methods in terms of relighting quality.

% \paragraph{Future work.}
There are still many potential future researching directions. Regarding the training time, combining our method with the popular 3D Gaussian splatting~\cite{kerbl20233d} may reduce the training time further. For the reconstruction quality, the efficient tensorial representation provides an opportunity to jointly update the geometry and material in the second stage, which may improve the reconstruction quality. From the perspective of diversity, reconstructing translucent objects is also worth researching. However, translucent materials need to consider higher dimensions, and designing an efficient formulation is the key. Our method currently focuses on single-object reconstruction with multi-view inputs. It would be interesting to apply our method on multi-object environments or few-shot reconstruction.

%% file: supp_content.tex
\section{Results}

\subsection{Reconstructed geometry evaluation}

We provide additional comparisons of the reconstructed geometry results on our synthetic dataset in Fig.~\ref{fig:geo_syn_comp_supp}. From the results, our method can not only reconstruct the surfaces with strong reflection in all scenes but also recover richer geometric details (\textsc{Compressor}, \textsc{Helmet} and \textsc{Dragon} scenes, etc.). On the \textsc{Motor} and \textsc{Dragon} scenes, our results are slightly higher than NeuS~\cite{wang2021neus} in MAE metric, mainly due to the noise problem, although our results are more detailed and closer to the ground truth visually.

\subsection{Estimated material evaluation}
We evaluate relighting quality on our synthetic dataset in terms of LPIPS~\cite{zhang2018lpips} in Tab.~\ref{tab:lpips}. Our method outperforms all the baseline methods. The visual results are shown in Fig.~\ref{fig:relight_syn_comp_supp}. By comparison, our method can produce decent reflection and highlights (such as \textsc{Compressor}, \textsc{Helmet} and \textsc{rover} scenes, etc.) based on the estimated materials, while the other methods either suffer from erroneous surfaces or tend to produce blurred reflections. 

% We also visualize the decomposition of estimated materials on \textsc{Helmet} and \textsc{Robot} scenes, as shown in Fig.~\ref{fig:mat_decomp}. The predicted albedo, metallic and roughness are very reasonable, leading to credible relighting results.

\begin{table}
  \centering
  \small
  \setlength\tabcolsep{2pt}
  \caption{Relighting quality in terms of LPIPS$\downarrow$ on our synthetic dataset. \textbf{Bold} means the best performance and \underline{underline} means the second best.}
  \label{tab:lpips}
  \begin{tabular}{ccccc}
    \hline
            & TensoIR & NeRO & NeILF++ & Ours \\
    \hline
    Rover          & 0.0801 & \underline{0.0693} & 0.0754 & \textbf{0.0593}  \\
    Dragon         & 0.1302 & \underline{0.0898} & 0.0988 & \textbf{0.0775}  \\
    Motor          & 0.0821 & \underline{0.0702} & 0.0870 & \textbf{0.0681}  \\
    Helmet         & \underline{0.1040} & 0.1079 & 0.1056 & \textbf{0.0770}   \\
    Robot          & 0.0931 & \underline{0.0755} & 0.0782 & \textbf{0.0613}    \\
    Compressor     & \underline{0.1038} & 0.1073 & 0.1286 & \textbf{0.0830}    \\
\hline
    Avg.           & 0.0989 & \underline{0.0867} & 0.0956 & \textbf{0.0710}    \\
\hline
\end{tabular}
\end{table}

\subsection{Results on real data}

We provide additional qualitative geometry and relighting results on the real data from NeILF++~\cite{zhang2023neilf++} in Fig.~\ref{fig:real_data_supp}. All results are reconstructed from LDR images. Our method is able to produce realistic relighting results based on our robust geometry and material reconstruction.

\subsection{Ablation studies on the shared tensor grid}
We adopt a shared tensor grid to encode geometry and appearance jointly in our TensoSDF representation, which can enhance the correlation between the geometry and appearance, while TensoRF~\cite{chen2022tensorf} uses two tensor grids to encode separately. We compare these two choices in Fig.~\ref{fig:tenso_ab}. From the results, using one shared tensor grid can improve the quality of reconstructed geometry.

% \paragraph{Gaussian smooth loss.}
% We introduce a Gaussian smooth loss to reduce the noise from tensorial representation. We show its impact in the second row of Fig.~\ref{fig:tenso_ab}, which can be seen that introducing the Gaussian smooth loss can reduce the noise on the reconstructed surface.

\myfigures{tenso_ab}{grid_ab}{The impact of the shared tensor grid. Compared to directly using two separate tensor grids to encode geometry and appearance respectively (left column), using one shared tensor grid can improve the quality of geometry reconstruction (right column).}{1.0\linewidth}

% \myfigures{mat_decomp}{mat_decomp2}{Decomposition of the estimated materials. Our method can produce reasonable albedo, metallic and roughness.}{1.0\linewidth}

% \mycfigure{nero_syn_comp_supp}{geo_nero_comp_supp.pdf}{Comparison of geometry reconstruction among our method, 
% TensoIR~\cite{jin2023tensoir}, NeILF++~\cite{zhang2023neilf++} and NeuS~\cite{wang2021neus} on the scene from NeRO~\cite{liu2023nero}. There is no available ground-truth mesh in NeRO datasets, and the CD$\downarrow$ loss is computed on the point cloud following NeRO~\cite{liu2023nero}.}

\mycfigure{geo_syn_comp_supp}{geo_comp_supp1.pdf}{Comparison of geometry reconstruction among our method, TensoIR~\cite{jin2023tensoir}, NeRO~\cite{liu2023nero}, NeILF++~\cite{zhang2023neilf++} and Neus~\cite{wang2021neus} on our synthetic dataset.}

\mycfigure{relight_syn_comp_supp}{relight_comp_supp.pdf}{Comparison of relighting results among our method, TensoIR~\cite{jin2023tensoir}, NeILF++~\cite{zhang2023neilf++} and NeRO~\cite{liu2023nero} on our synthetic dataset.}

\mycfigure{real_data_supp}{real_data_supp.pdf}{Our reconstructed geometry and relighting results on the real data from NeILF++~\cite{zhang2023neilf++}. All results are reconstructed from LDR images.}